\documentclass[12pt]{article}
\usepackage{authblk}
\usepackage[bookmarksnumbered, colorlinks, plainpages]{hyperref}
\usepackage{amsmath, amsthm, amscd, amsfonts, amssymb, graphicx, color, booktabs}
\newcommand{\kl}{K_{\rm L}}
\newcommand{\klmm}{K_{\rm L}\rightarrow\mu^+\mu^-}
\newcommand{\gf}{G_{\rm F}}
\newcommand{\chw}{\mathcal{H}_{\rm W}}
\newcommand{\tsep}{t_{\rm sep}}
\newcommand{\lla}{\left\langle}
\newcommand{\rra}{\right\rangle}

\numberwithin{equation}{section}
\definecolor{email}{rgb}{0.00,0.00,0.84}
\begin{document}
\setcounter{page}{1}

\title{\large \bf 12th Workshop on the CKM Unitarity Triangle\\ Santiago de Compostela, 18-22 September 2023 \\ \vspace{0.3cm}
\LARGE Progress on $\klmm$ from Lattice QCD}

\author[1]{En-Hung Chao}
\affil[1]{Department of Physics, Columbia University, New York, NY 10027, USA \\
\{en-hung.chao\}@m4x.org
}

\maketitle

\begin{abstract}
In this contribution, we present recent progress from the RBC/UKQCD collaboration on the first calculation of the long-distance two-photon contribution to the decay amplitude of a long-lived kaon into a pair of charged muons. 
\end{abstract} \maketitle

\section{Introduction}
\begin{figure}
\centering
\includegraphics[scale=0.3]{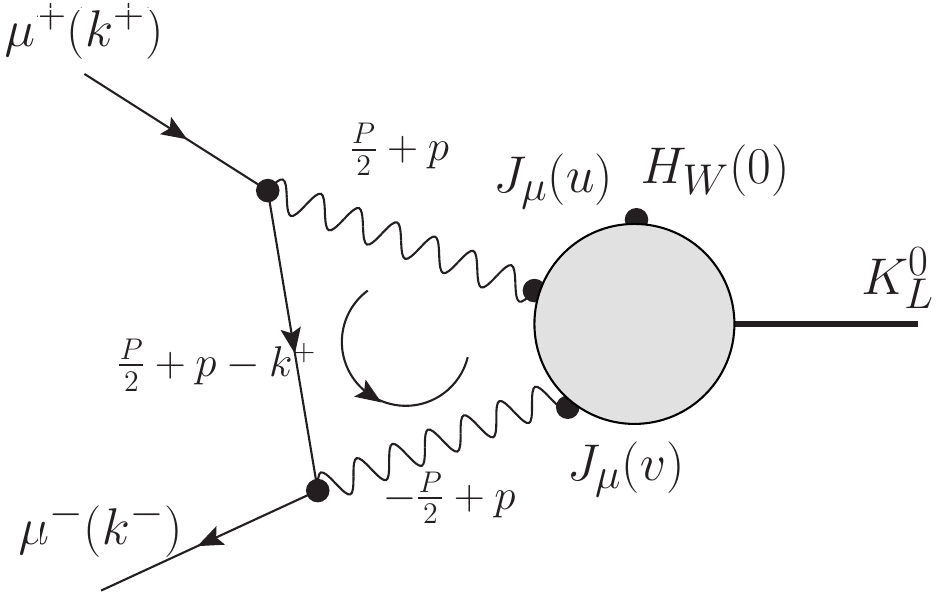}
\caption{The two-photon long-distance contribution to the $\klmm$ decay.}
\label{fig:2gLD}
\end{figure}

The $\klmm$ decay is considered as one of the golden rare kaon decay modes for its experimental cleanness. 
The experimental value for the branching ratio, $\textrm{Br}(\klmm)=6.84(11)\times 10^{-9}$, has reached an accuracy of $1.6\%$ two decades ago~\cite{E871:2000wvm}, and could potentially serve as a precision probe for the Standard Model (SM).
The short-distance part of the decay amplitude due to exchange of $W$- and $Z$-bosons has been determined to a $15\%$-precision in the SM~\cite{Gorbahn:2006bm}.
However, to make a meaningful comparison to experiment, one also needs a precise determination of the long-distance (LD) contributions to the amplitude, among which the process exchanging two photons (Fig.~\ref{fig:2gLD}) is known to contribute considerably to the absorptive part.
While the latter can be reconstructed from other experimentally-measurable processes based on optical theorem, computing the dispersive part non-perturbatively is much more challenging.
In this work, we present a formalism which allows the first-principles computation of both the dispersive and the absorptive parts of the two-photon LD contribution to the decay amplitude from lattice QCD~\cite{Christ:2020bzb, Christ:2020dae, Zhao:2022pbs, Christ:2022rho, Chao:2023cxp}.
Part of these proceedings overlaps with materials presented at the 2023 Lattice Symposium~\cite{Chao:2023cxp}.
We refer the reader to those conference proceedings for a more exhaustive list of references. 
 
\section{Formalism}
Inspired by the successful calculations of the hadronic light-by-light scattering contribution to the anomalous magnetic moment of the muon from lattice QCD~\cite{Blum:2019ugy,Chao:2021tvp}, we express the decay amplitude with quantities in coordinate space.
We perform a perturbative expansion to O($\alpha^2$) in the QED sector and use an effective weak Hamiltonian $\chw$ run down to an energy scale below the charm-quark mass at O($\gf$), where $\alpha$ is the fine-structure constant and $\gf$ is the Fermi decay constant.
Starting from the Minkowski-space expression for the $\kl\rightarrow\mu^+\mu^-$ decay amplitude (cf. Fig.~\ref{fig:2gLD})
\begin{equation}\label{eq:amp}
\mathcal{A}(k^+,k^-) = 16\pi^2\alpha^2 \; \int\!\! d^4 u \; \int\!\! d^4 v \; 
\mathcal{L}_{\mu\nu}(u-v) \left\langle 0\left|\textrm{T}\left\{ J_\mu(u) J_\nu(v) \mathcal{H}_W(0)\right\}\right|K_L\right\rangle\,,
\end{equation}
where $\mathcal{L}_{\mu\nu}$ is a QED kernel and the $J_\lambda$'s are electromagnetic (EM) currents, we Wick-rotate the coordinate-space variables $u$ and $v$ so that the hadronic matrix element can be evaluated in Euclidean space on the lattice.
However, after analytic continuation, the kernel grows exponentially at large temporal separations, leading to unphysical, exponentially growing contributions to the right-hand side of Eq.~\eqref{eq:amp}.
Fortunately, most of the intermediate states which can be inserted between the operators are sufficiently massive to introduce stronger exponential suppressions at large time separations to compensate the exponential growth of the kernel. 
With the physical quark masses, the sources of unsuppressed unphysical exponential growth in the analytically-continued amplitude are:
\begin{enumerate}
\item[S1:] $\pi^0$ with zero spatial momentum, created from the $\kl$ by $\chw$;
\item[S2:] Slow-moving $\pi\pi$ states propagating between the two EM currents with an energy $2m_\pi < E_{\pi\pi}\leq M_K$\footnote{These are $\pi\pi\gamma$ states with $2m_\pi\leq E_{\pi\pi\gamma}\leq M_K$ if the photon contained in $\mathcal{L}_{\mu\nu}$ is also identified.}.
\end{enumerate}
To extract the physical amplitude, one can compute Eq.~\eqref{eq:amp} in Euclidean space regulated in the IR by a finite temporal integration upper bound and subtract the calculable unphysical contributions.

\section{Numerical implementation}
Our first goal is to apply the formalism to the 24ID physical-pion-mass ensemble from the RBC/UKQCD collaboration.
The parameters of this ensemble are given in Tab.~\ref{tab:24id}.
In order to determine Eq.~\eqref{eq:amp}, we calculate the quantity on the lattice:
\begin{equation}\label{eq:master}
\mathcal{A}(\tsep, \delta_{\rm max}, x) \equiv \sum_{u,v\in\Lambda, v_0\le u_0\atop v_0-x_0\le \delta_{\rm max}} e^{M_K (v_0-t_K)}\ \mathcal{L}_{\mu\nu}(u-v)\lla\textrm{T}\{ J_\mu(u) J_\nu(v)\chw(x)\kl(t_K)\}\rra\,.
\end{equation}
In the above, $\tsep\equiv x_0-t_K$ needs to be taken large in order to project onto the $\kl$ ground state.
The exponential factor here guarantees that the expression gives the same answer at large $\tsep$ up to statistical fluctuations.
In practice, we exploit translational invariance to average Eq.~\eqref{eq:master} computed at different fixed reference points $x$ to reduce the statistical error.
The upper bound $\delta_{\rm max}$ limits the time through which the intermediate $\pi^0$ state can propagate, allowing the exponential growth caused by this state to be studied and its removal verified.
For this work, only two of the $\Delta S=1$ operators are considered~\cite{Buchalla:1995vs}
\begin{equation*}\small
\chw(x) = \frac{G_{\rm F}}{\sqrt{2}}V_{us}^*V_{ud}(C_1Q_1+C_2Q_2)\,,
\end{equation*}
where the $V_{ab}$'s are CKM matrix elements and $C_{1,2}$ the Wilson coefficients obtained from a non-perturbative matching procedure~\cite{Zhao:2022njd}.

\begin{table}[h!]
\centering
\begin{tabular}{c c c c}
\toprule
$L^3\times T\times L_s$ & $m_\pi$ [MeV] & $M_K$ [Mev] & $a^{-1}$ [GeV] \\
\midrule
$24^3\times 64\times 24$ & 142 & 515 & 1.023\\
\bottomrule
\end{tabular}
\caption{Parameters of the 24ID ensemble.}
\label{tab:24id}
\end{table}

With this ensemble, the first excited $\pi\pi\gamma$ state is expected to be already more energetic than the kaon; therefore, we anticipate that there will be no unphysical divergence due to the states (S2).
As a consequence, the only unphysical contribution to be subtracted from Eq.~\eqref{eq:master} is that from the zero-momentum $\pi^0$ (S1), given by:
\begin{equation}\label{eq:pi0}
\begin{split}
\frac{1}{2m_\pi}\sum_{u,v\in\Lambda,v_0\le u_0\atop 0\leq v_0-x_0\leq \delta_{\max}}e^{(M_K-m_\pi)(v_0-x_0)}
\lla 0| J_\mu(u)J_\nu(v)|\pi^0\rra \mathcal{L}_{\mu\nu}(u-v)
\lla\pi^0|\chw(v)|\kl\rra\,,
\end{split}
\end{equation}
up to a normalization constant.
To verify that the $\pi\pi\gamma$-intermediate states do not introduce exponential growth, we perform calculations at different cutoffs $R_{\rm{max}}$ on the temporal separation of the two EM currents, ie. $|u_0-v_0|\leq R_{\rm{max}}$.

\begin{figure}[h!]
\centering
\includegraphics[scale=0.55]{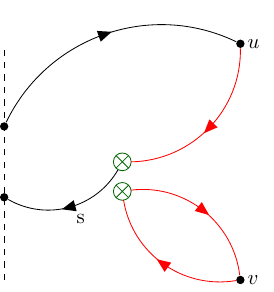}
\includegraphics[scale=0.55]{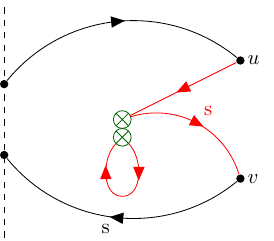}
\includegraphics[scale=0.55]{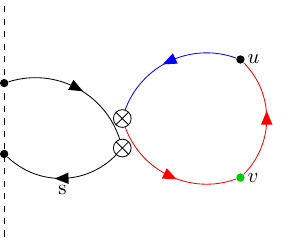}
\includegraphics[scale=0.55]{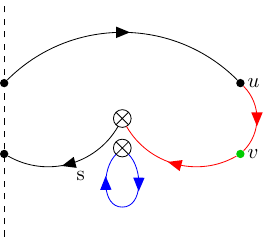}
\includegraphics[scale=0.55]{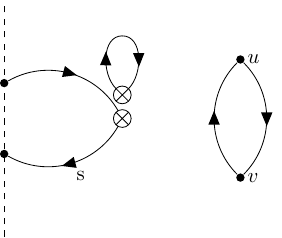}
\caption{Different Wick-contraction topologies. From left to right: Type 1, 2, 3, 4 (connected) and 5 (disconnected). The dashed lines correspond to the $\kl$ interpolator, the filled dots to the EM currents and the crosses to $\chw$.}
\label{fig:contr-diag}
\end{figure}

\section{Preliminary results on the connected diagrams}
\begin{figure}[h!]
\centering
\includegraphics[scale=0.3]{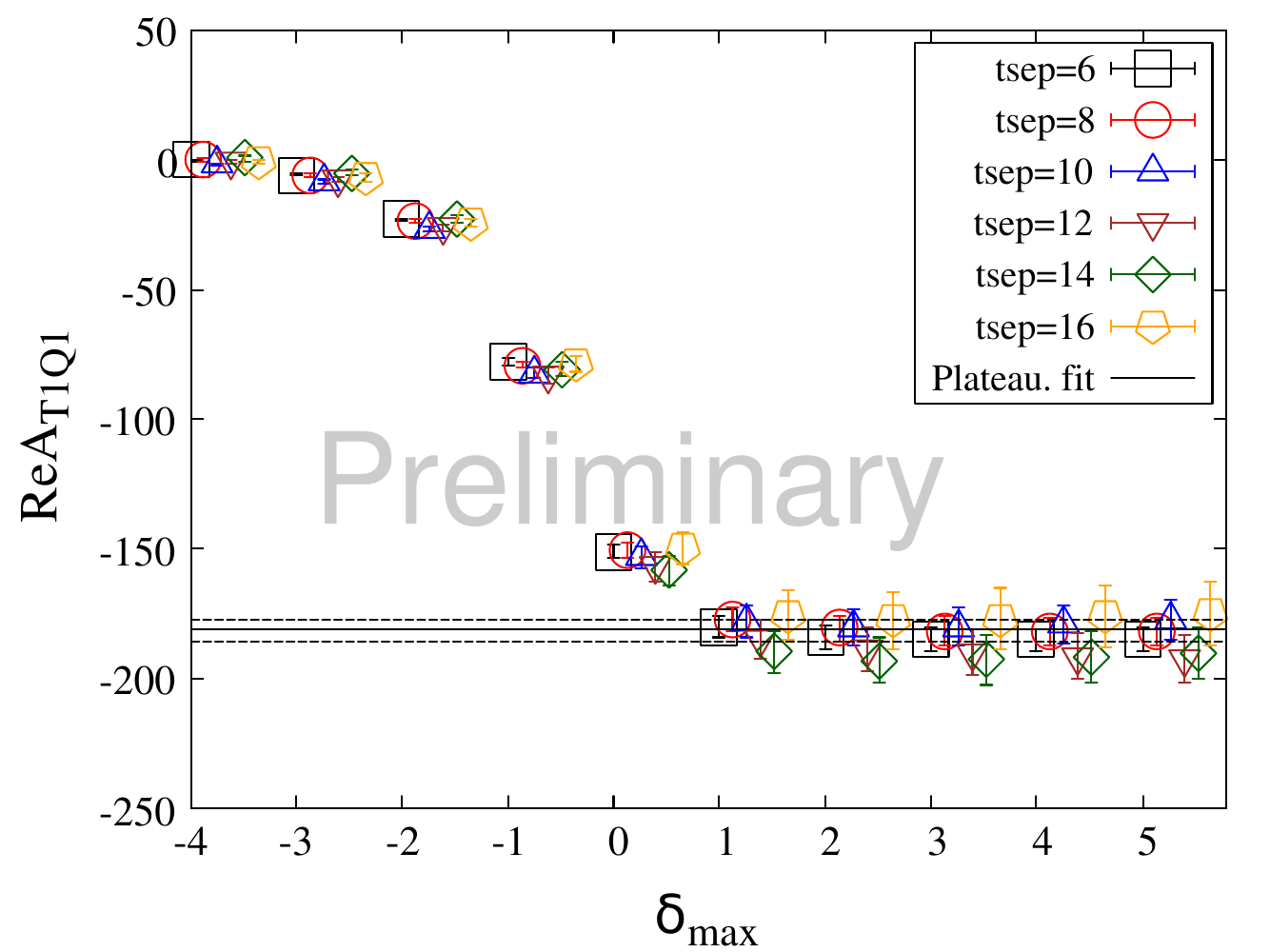}
\includegraphics[scale=0.3]{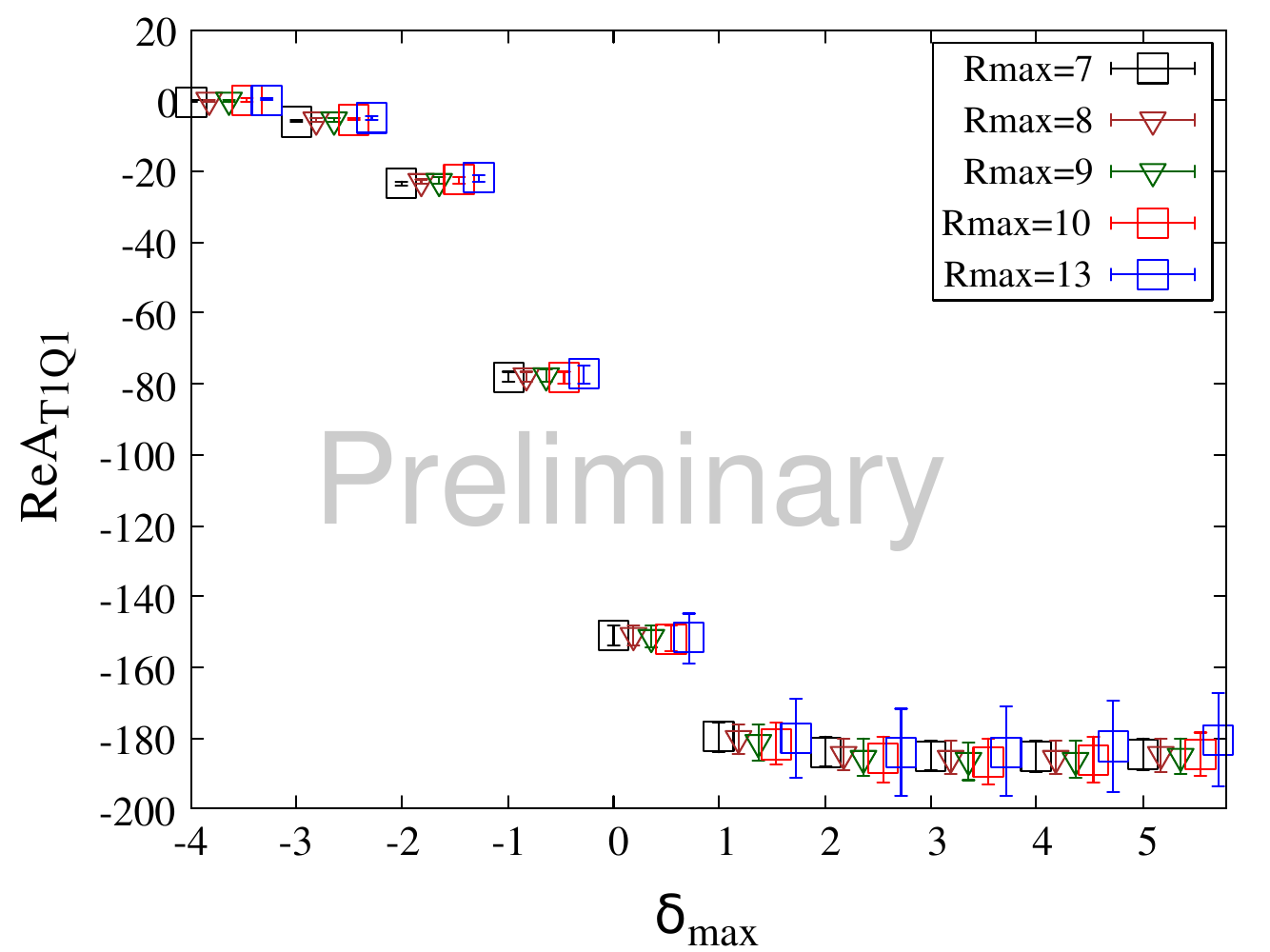}
\caption{Left: $\mathcal{A}_{\rm T1Q1}$ at fixed $R_{\rm max}=7$. Right: $\mathcal{A}_{\rm T1Q1}$ at fixed $\tsep=6$}\label{fig:t1q1}
\end{figure}

\begin{figure}[h!]
\centering
\includegraphics[scale=0.3]{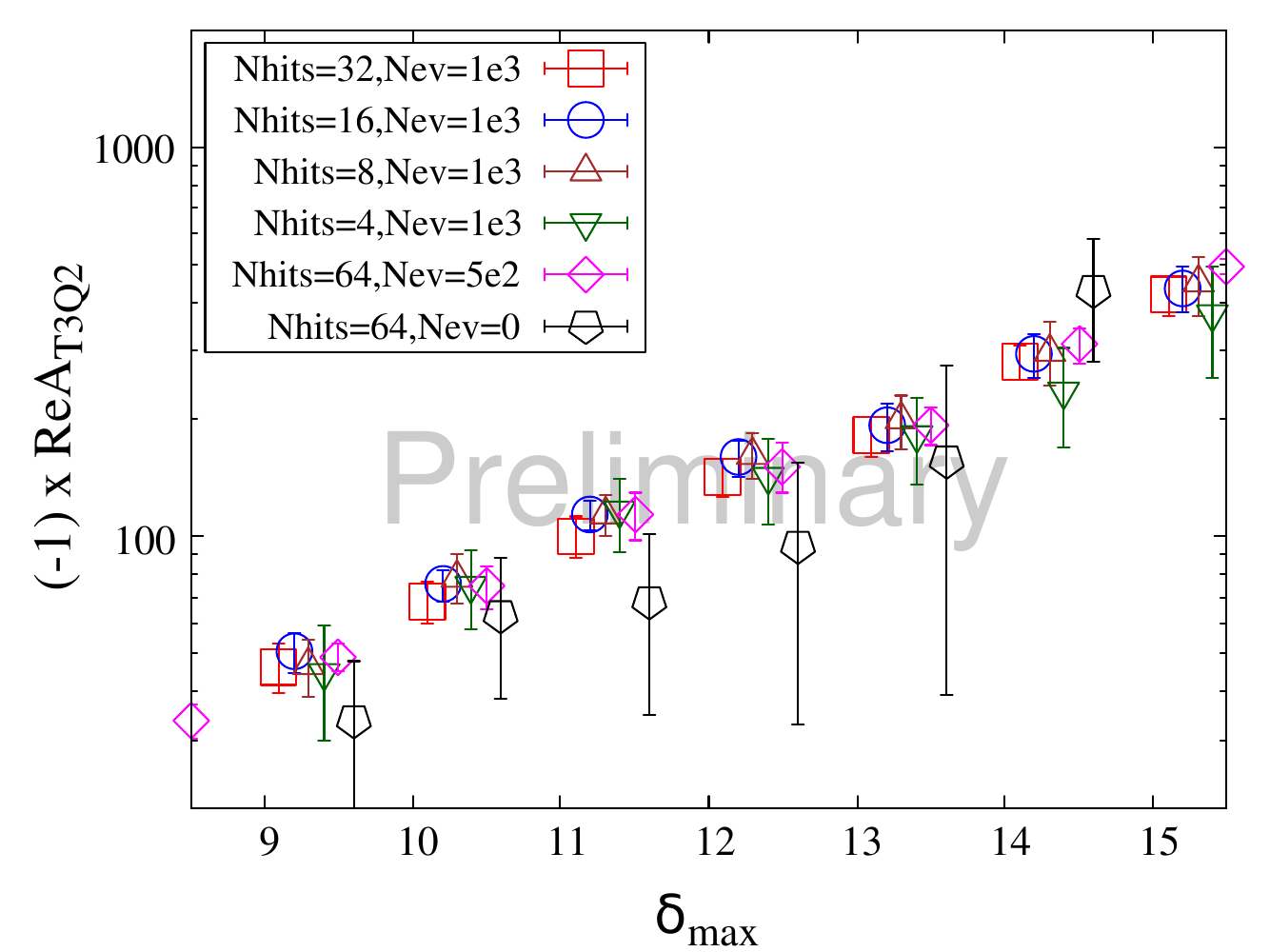}
\includegraphics[scale=0.3]{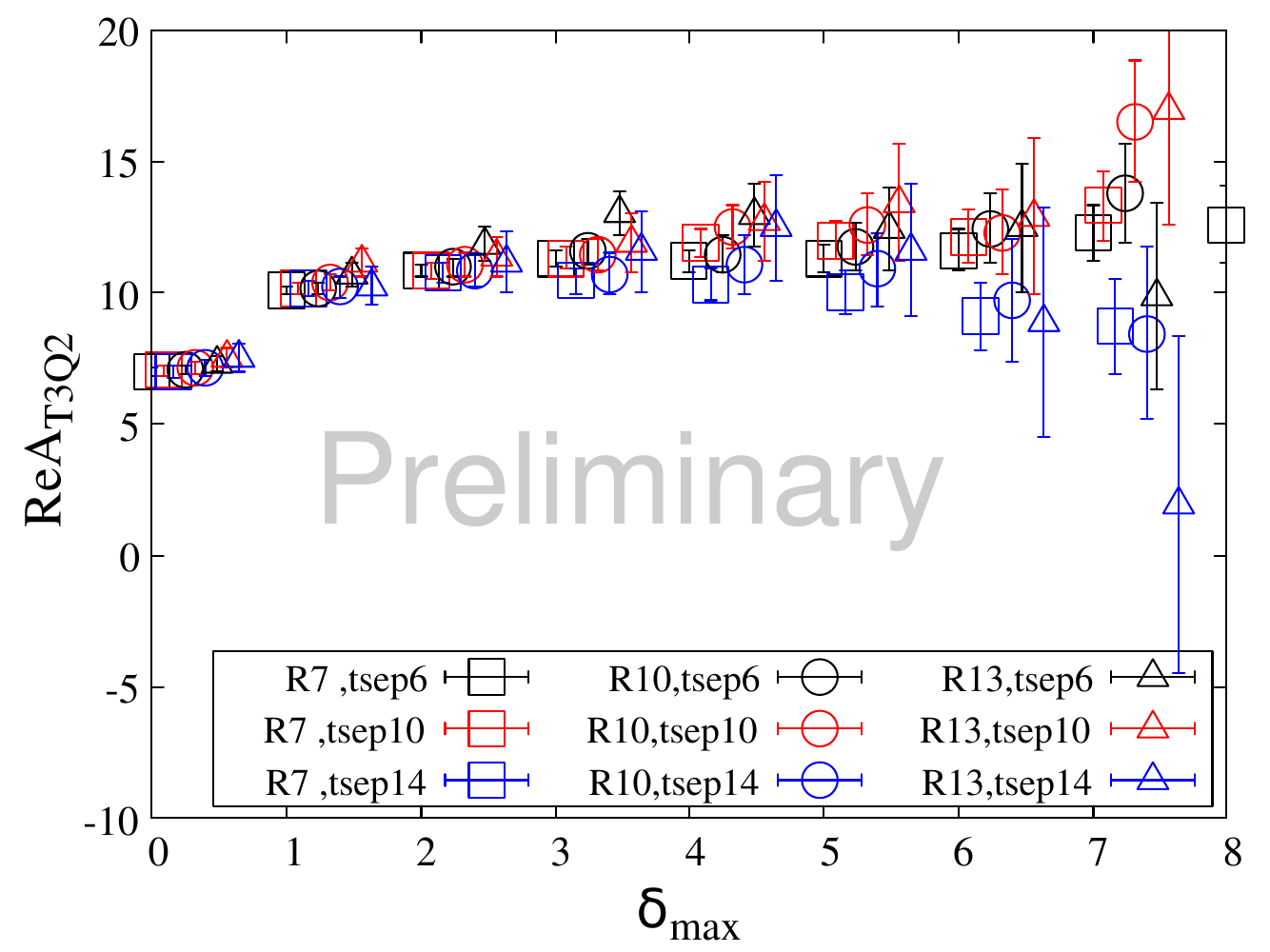}
\caption{Left: $\mathcal{A}_{\rm T3Q2}$ in log-scale with different hits and low modes. The displayed results are computed with the all-to-all propagator method~\cite{Foley:2005ac}, where `Nhits' and `Nev' correspond to the number of stochastic sources and low-modes used. Right: $\mathcal{A}_{\rm T3Q2}$ after subtracting $\pi^0$}\label{fig:t3q2}
\end{figure}

We first compute the quark-connected contributions as they are expected to have less statistical fluctuations than their disconnected counterparts (see Fig.~\ref{fig:contr-diag}).
A series of strategies to efficiently evaluate Eq.~\eqref{eq:master} for these contraction topologies at several values of $\tsep$ and $R_{\rm max}$ have been established and implemented. 
Only for the time-ordering $\chw$ is earlier than the two EM currents can an unphysical $\pi^0$ state occur, as can be seen in Fig.~\ref{fig:contr-diag}.
In this case, it is possible to create a $\pi^0$ only if there are no more than two quark lines between $\chw$ and the two EM currents at a given time.
Therefore, only the Type-3 and -4 diagrams can produce an unphysical $\pi^0$ among the connected contributions. 
We will comment on selected results for the Type-1 and Type-3 diagrams for the real part of Eq.~\eqref{eq:master}, displayed in Fig.~\ref{fig:t1q1} and Fig.~\ref{fig:t3q2}.
Similar features are observed for the Type-2 and Type-4 diagrams.
A more detailed description of the computational setup can be found in Ref.~\cite{Chao:2023cxp} and is the subject of a forth-coming publication. 

The Type-1 results in the left panel of Fig.~\ref{fig:t1q1} reach a plateau in $\delta_{\rm max}$ rapidly as expected, due to the absence of the unphysical $\pi^0$.
The good agreement between results from different $\tsep$'s suggests that the $\kl$ ground state is correctly produced and the results at different $\tsep$ can be averaged over to further reduce the statistical error.
In the right panel, we show results with different $R_{\rm max}$.
The central values of these results lie consistently on top of each other, which confirms that there is no unphysical $\pi\pi\gamma$-state contribution in the allowed kinematics.

The Type-3 results in the left panel of Fig.~\ref{fig:t3q2} exhibit a clear exponential growth due to the unphysical intermediate $\pi^0$ contribution. 
After subtracting the quantity given in Eq.~\eqref{eq:pi0} from the data, we can identify a clear plateau formed from rather small $\delta_{\rm max}$ across different $R_{\rm max}$ and $\tsep$ values (right panel of Fig.~\ref{fig:t3q2}), luckily before the signal deteriorates quickly as we move to larger $\delta_{\rm max}$.

\section{Recent developments on the disconnected part}
\begin{figure}
\centering
\includegraphics[scale=0.3]{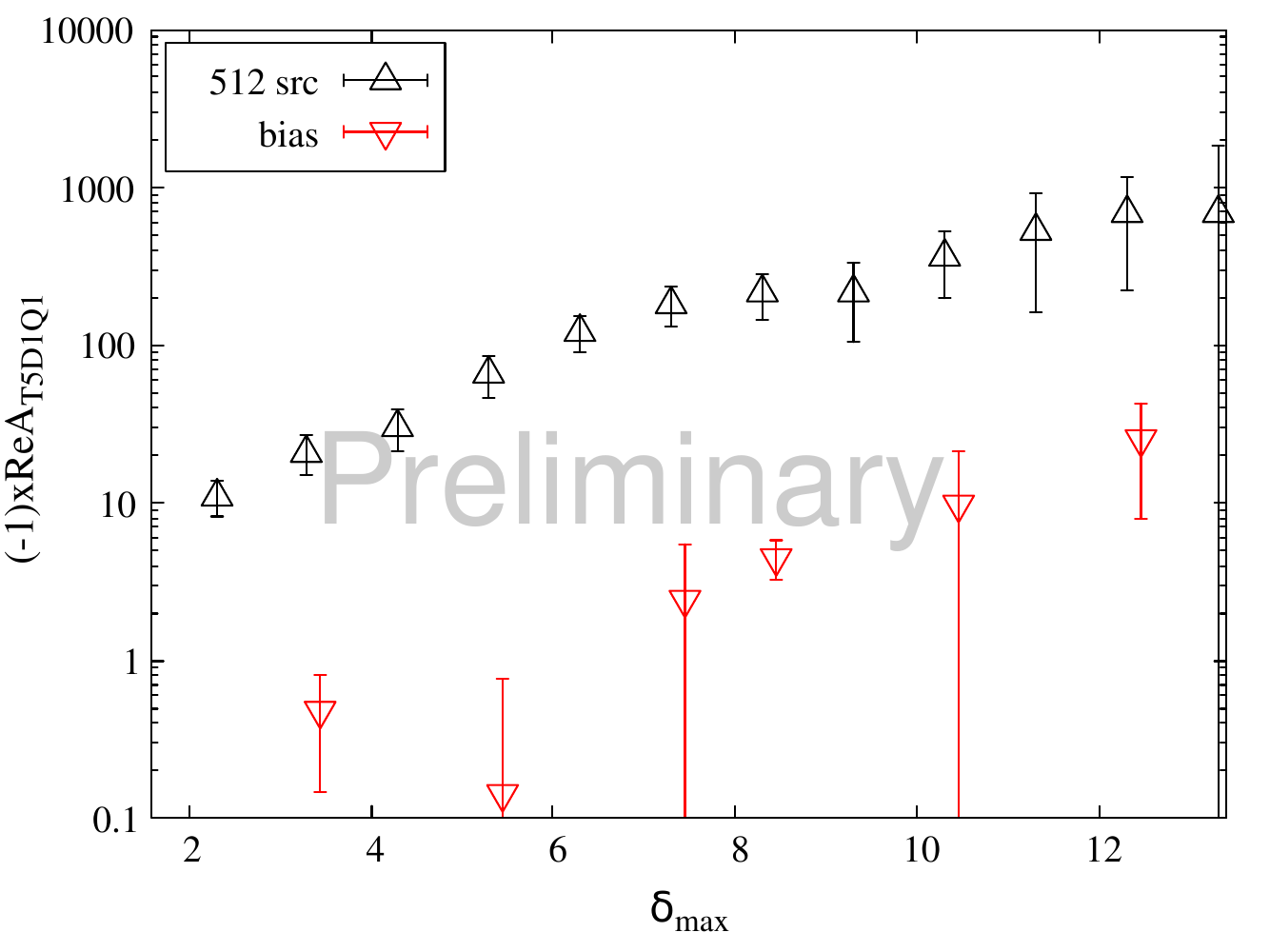}
\includegraphics[scale=0.3]{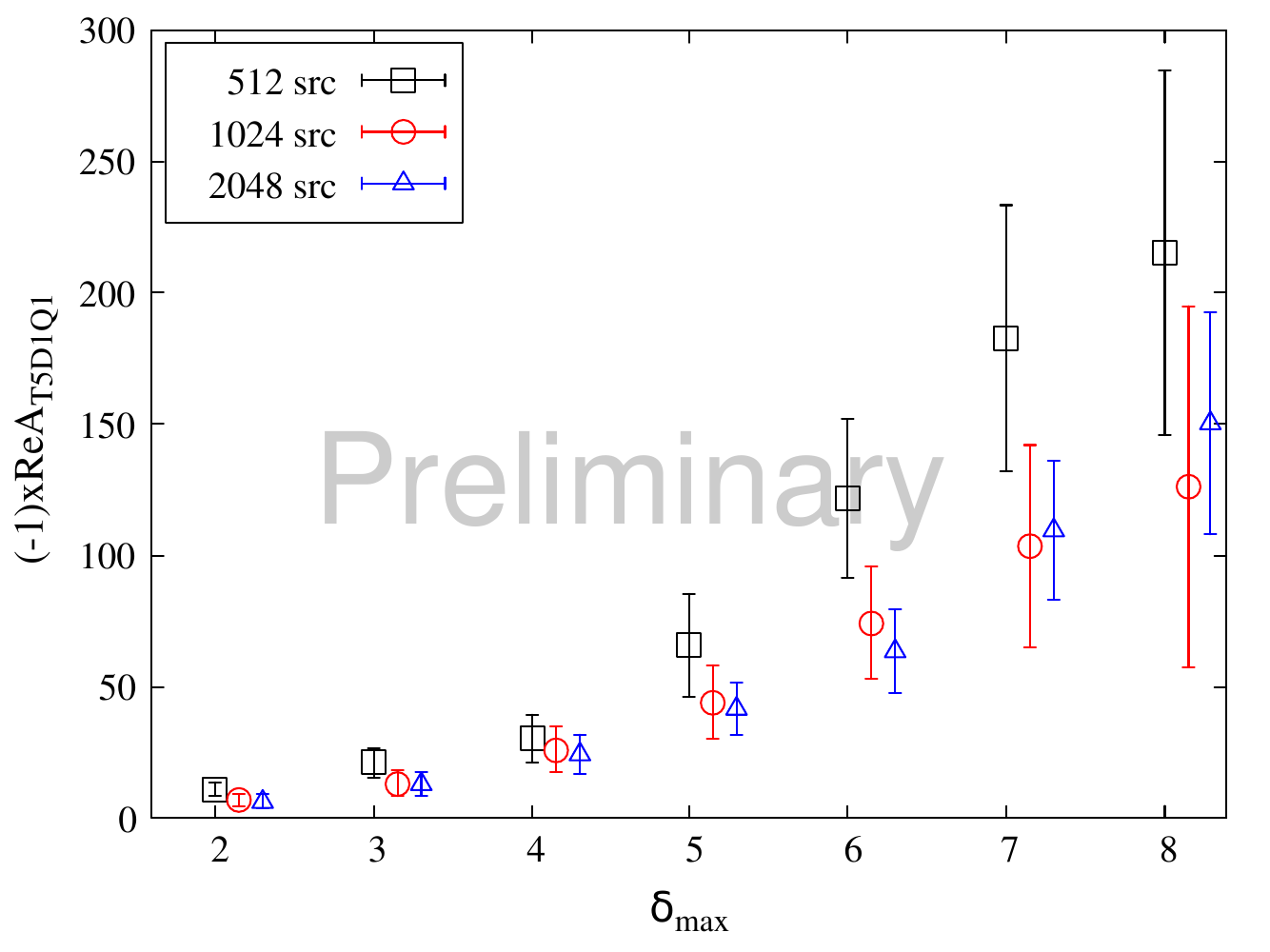}
\caption{Left: bias of the data computed from propagators with looser stopping condition compared to the original data on 512 source points. Right: Comparison between data computed on different numbers of source points.}
\label{fig:type5}
\end{figure}

Since CKM2023, there has been some progress worth reporting on our attempt to reduce the statistical noise on the quark-disconnected part. 
Our sampling strategy requires swapping the r\^oles of $x$ and $v$ in Eq.~\eqref{eq:master} using translational invariance to make the latter our new reference point.
In our first series of calculations, the disconnected part was constructed with 512 existing point-source propagators, each with a distinct reference point.
The statistical noise was about an order of magnitude larger than that from the connected diagrams with the same number of configurations averaged over the same number of reference points. 
To bring it to a comparable level of statistical noise, we have been putting effort into accumulating data on more reference points, but with propagators which are partly obtained by solving the Dirac equation with a low-precision conjugate gradient (CG) inversion.
With the latter, we can increase the statistics and thus reduce the error with significantly less computing time providing that the fluctuation of the incurred bias compared to utilizing higher precision solves is negligible compared to the statistical noise~\cite{Bali:2009hu,Blum:2012uh}.
Our preliminary study displayed in the left panel of Fig.~\ref{fig:type5} shows that the bias fluctuates about an order of magnitude less than the data with propagators computed with high-precision CG in the relevant region of $\delta_{\rm max}$.
In the right panel of Fig.~\ref{fig:type5}, we can see that the error on the averaged data is halved as we quadruple the number of reference points, suggesting that this is an effective error reduction scheme.
Typically, we observe that the computing time to reach a CG-residue of $10^{-4}$ is about three-times less than needed for a CG-residue of $10^{-10}$ on the 24ID ensemble.

\section{Summary and outlook}
In this contribution, we summarize the ongoing effort on calculating the long-distance contribution to the $\klmm$ decay amplitude from lattice QCD using a coordinate-space based formalism.
In particular, this important amplitude, which must be known if we are to compare the second-order Standard Model weak interaction prediction with experiment, can be determined from first principles with our method.
The difficulties arising from analytically continuing the amplitude from Minkowski space to Euclidean space are addressed with well-established numerical strategies for the quark-connected part.
Following the achievable precision, we expect the disconnected part to dominate the total error budget.
Since the CKM2023 Workshop, promising error-reduction strategies to deal with this more challenging quark-disconnected part have been developed.


{\bf Acknowledgements.} We thank our colleagues from the RBC/UKQCD collaboration for useful discussions and substantial technical support. This work is supported by the U.S. D.O.E. grant \#DE-SC0011941.

\bibliographystyle{amsplain}

\end{document}